\documentclass[doublespacing]{elsart}

\usepackage[dvips]{graphicx,color}
\def\disfrac#1#2{{\displaystyle\frac{#1}{#2}}}
\def\eqn#1{{\begin{eqnarray}#1\end{eqnarray}}}
\def\eq#1{{\begin{eqnarray*}#1\end{eqnarray*}}}
\newcommand{\mitDelta}{{\it \Delta}}

\usepackage{amssymb}

\begin{document}

\begin{frontmatter}

\title{Ehrenfest Model with Large Jumps in Finance}
\author[label1]{Hisanao TAKAHASHI\thanksref{previous}}
\thanks [previous]{Previous address: Department of Statistical Science, School of Mathematical and Physical Science,
The Graduate University for Advanced Studies (Sokendai).
4-6-7 Minami-Azabu, Minato-ku, Tokyo 106-8569, Japan}
\address[label1]{The Institute of Statistical Mathematics, 4-6-7 Minami-Azabu, Minato-ku, Tokyo 106-8569, Japan}
\ead{hisanao@ism.ac.jp}
\date{25 July 2003}

\begin{abstract}
Changes (returns) in stock index prices and exchange rates for currencies are argued, based on empirical data, to obey a stable distribution with characteristic exponent $ \alpha < 2 $ for short sampling intervals and a Gaussian distribution for long sampling intervals.
In order to explain this phenomenon, an Ehrenfest model with large jumps (ELJ) is introduced to explain the empirical density function of price changes for both short and long sampling intervals.
\end{abstract}

\begin{keyword}
L\'evy stable process; Ehrenfest model; Finance
\PACS 02.50.-r; 05.45.Tp; 89.65.Gh
\end{keyword}

\end{frontmatter}

\section{Introduction}

This paper introduces a simple stochastic model, an Ehrenfest model with large jumps (ELJ), to explain the observed fact that the density function of price change fits well to a symmetrical L\'evy stable distribution near the central part of the density function, while it has a truncated tail far from the center \cite{Stanley}.
The proposed model also explains that it is close to a Gaussian distribution for long sampling intervals.

Changes in stock prices and exchange rates for currencies are one of the best examples to which the random walk concept is applicable \cite{Hull}.
We have seen a lot of work of related topics since the work of L. Bachelier in 1900.
In financial theory such as the argument on the Black-Scholes model, the distribution of (logarithmic) stock returns is assumed to be Gaussian.
It seems that this assumption is well accepted for a long sampling interval of empirical data, namely those for sampling intervals of several days, a week or longer.

In the case of a short sampling interval of empirical data, i.e. those for sampling intervals of several seconds, minutes or hours, the price returns are not distributed according to a Gaussian law \cite{Barndorff-Nielsen}.
The value of $ \alpha $ for the empirical density function of the S\&P500 index estimated for short sampling intervals indicates that the density function follows a stable law ($ \alpha < 2 $) \cite{Stanley,Stanley 2}.

The price changes are therefore said to be distributed according to a stable law with a characteristic exponent $ \alpha < 2 $ for short sampling intervals, and a Gaussian distribution ($ \alpha = 2 $) for long sampling intervals \cite{Eberlein 1995}.

\section{Ehrenfest model with large jumps}

The Ehrenfest model \cite{Aoki,Feller} envisages two boxes $ + $ and $ - $, with $ 2R $ particles distributed in these boxes.
A particle is chosen at random and moved from one box to the other and the same procedure is repeated.

We consider a generalized Ehrenfest model in which $ a\,b^j $ steps of the Ehrenfest model take place with probability $ C/Q^{j} $
at each step,
where $ a,\ b>0,\ Q>1,\ j=0,\ 1,\ 2,... $ and $ C $ is a normalization constant defined by:
\eq{
	C= 1-\disfrac{1}{Q}.
}%
For simplicity, $ a $ is fixed to be 1 in the following discussion.
This generalized model becomes identical to the original Ehrenfest model in the limit $ Q \to \infty $.

Suppose that initially there are $ R+n $ ($ -R\le  n\le R $) particles in box $ + $, after repeating the above procedure $ s $ times, there are $ R+m $ particles in box $ + $.
The probability of this event calculated in case of the original Ehrenfest model \cite{Kac} can be described as:
\eqn{
	P(n|m;s)
	=
	\frac{(-1)^{R+j}}{2^{2R}}\sum_{j=-R}^R \left(\frac{j}{R}\right)^s C_{R+j}^{(-n)}C_{R+m}^{(j)},
	\label{Ehrenfest Kac}
}%
where
\eq{
	(1-z)^{R-j}(1+z)^{R+j}\equiv \sum_{k=0}^{2R} C_k^{(j)} z^k.
}%

With the duration of each step for the original Ehrenfest model defined as $ \tau $, consider a particle moving along the $ x $-axis in such a way that at the time $ s\, \tau $ it is located at $ \Delta \,m $.
In the diffusion limit:
\eq{
	\Delta\to 0,\quad \tau\to 0,\quad \frac{\Delta^2}{2\tau}=D,
	\quad \frac{1}{R\tau}\to\gamma,\quad s\tau=t,\quad m\Delta\to x,
}%
we have
\eq{
	\lim \sum_{x_1<m\Delta <x_2}P(n|m;s)=\int_{x_1}^{x_2}P(x_0|x;t)\,dt,
}%
where
\eqn{
	P(x_0|x; t)
	=
	\sqrt{\frac{\gamma}{2\pi D(1-e^{-2\gamma t})}}
	\exp\Bigg[-\frac{\gamma(x-x_0e^{-\gamma t})^2}{2D(1-e^{-2\gamma t})}\Bigg].
	\label{Ehrenfest P(x_0|x; t)}
}%

We denote the particle number in the box $ + $ after $ s $ steps as $ M(s)$, and define the changes as:
\eq{
	Z_{\mitDelta s}(s) = M(s)-M(s-\mitDelta s),
}
where $ \mitDelta s $ denotes the step interval.
Using the above $ Z_{\mitDelta s}(s) $, we define the empirical density $ P_{\mitDelta s}(z) $ as:
\eqn{
	\sum_{z_1<z<z_2} P_{\mitDelta s}(z)=
	\sum_{z_1<z<z_2} \frac{\displaystyle\sum_s \delta_{z,Z_{\mitDelta s}(s)}}{s_{\rm last}},
	\label{the empirical density}
}%
where $ s = 1, 2, 3,...,s_{\rm last} $ and $ \delta_{l,k} $ denotes the Kronecker delta.

Using Eq. (\ref{Ehrenfest Kac}), the probability density function of the changes $ Z_{\mitDelta s}(s) $ for the proposed model $ P^{\rm ELJ}_{\mitDelta s}(z) $ is expressed as:
\eqn{
	&&
	P^{\rm ELJ}_{\mitDelta s}(z)
	\nonumber
	\\
	&=&
	\sum_{m,n}\ \sum_{i_1,i_2,...,i_{\mitDelta s}}\ \sum_{n_1,n_2,...,n_{\mitDelta s}}
	\frac{C}{Q^{i_1}}P(n|n_1;b^{i_1})\frac{C}{Q^{i_2}}P(n_1|n_2;b^{i_2}) \times \cdots
	\nonumber
	\\
	&&\hspace{13em}
	\times
	\frac{C}{Q^{i_{\mitDelta s}}}P(n_{\mitDelta s}|m;b^{i_{\mitDelta s}}) \delta_{m-n,k},
	\label{P(z) Our Kac}
}%
where $ i_k $ = 0,1,2,....
Using the diffusion approximation, the above equation can be approximated by Eq. (\ref{P(z) Our Ehrenfest}).

Assuming that the initial density function is given by a Gaussian density function $ g(x) = P(0|x; \infty)$, the probability density function of the change $ Z_{\mitDelta t} (t) $ in the original Ehrenfest model, obtained from the diffusion approximation (\ref{Ehrenfest P(x_0|x; t)}), is given by:
\eqn{
	&&
	P^{\rm Ehr.}_{\mitDelta t}(z)
	\nonumber
	\\
	&=&
	\int P(x_0|x;\mitDelta t)\,g(x_0)\,\delta(z-(x-x_0))\, dx\,dx_0
	\nonumber
	\\
	&=&
	\frac{\gamma}{2\pi D\sqrt{1-e^{-2\gamma t}}}
	\int
	\exp\Bigg[-\frac{\gamma(z+x_0-x_0e^{-\gamma t})^2}{2D(1-e^{-2\gamma t})}\Bigg]
	\exp\Bigg[-\frac{\gamma\,x_0^2}{2D}\Bigg]\,dx_0
	\nonumber
	\\
	&=&
	\sqrt{\frac{\gamma}{4\pi D (1-e^{-\gamma\,\mitDelta t})}}
	\exp
	\left(-\frac{\gamma}{4D}
		\frac{z^2}{1-e^{-\gamma\,\mitDelta t}}\right),
	\label{P(z) Ehrenfest}
}%
where $ \delta(x) $ is the Dirac delta function.
The variance of the change $ Z_{\mitDelta t} (t) $ is given by:
\eq{
	{\rm Var}(Z_{\mitDelta t} (t))
	=
	\frac{2D}{\gamma}({1-e^{-\gamma\,\mitDelta t}}).
}%

\begin{figure}
	\begin{center}
	\rotatebox{90}{\hspace{2.5cm}$ f(t) $}
	\includegraphics[height=6cm]{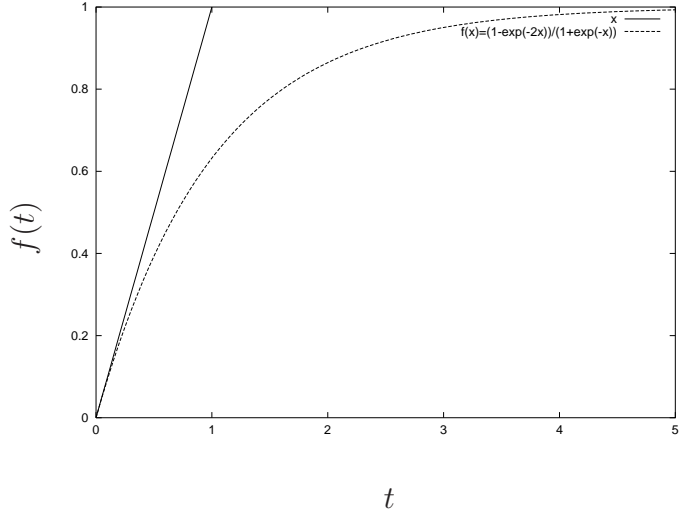}\\
	\qquad~$ t $
	\caption[]{Graph of function $ f(t) = (1-e^{- t}). $}
	\label{fig. variance}
	\end{center}
\end{figure}

Figure \ref{fig. variance} shows the behavior of the function:
\eq{
	f(t)={1-e^{- t}}.
}%
From this figure, we find that when both $ \mitDelta t_1 $ and $ \mitDelta t_2 $ are sufficiently large, the difference $ {\rm Var}(Z_{\mitDelta t_1} (t))-{\rm Var}(Z_{\mitDelta t_2} (t)) $ remains small, even if $ \mitDelta t_1 $ and $ \mitDelta t_2 $ are very different.

The time series of the ELJ is interpreted as a time series chosen randomly from that of the original Ehrenfest model.
Using Eq. (\ref{P(z) Ehrenfest}), the probability density function of the ELJ is approximated as:
\eqn{
	& &
	P^{\rm ELJ}_{\tau\mitDelta s}(z)=
	\sum_{i_1,...,i_{\mitDelta s}} \frac{C^{\mitDelta s}}{Q^{i_1+\cdots+i_{\mitDelta s}}}P^{\rm Ehr.}_{\tau ( b^{i_1}+\cdots+b^{i_{\mitDelta s}})}(z)
	\nonumber
	\\
	&=&
	\sum_{i_1,...,i_{\mitDelta s}} \frac{C^{\mitDelta s}}{Q^{i_1+\cdots+i_{\mitDelta s}}}
	\sqrt{\frac{\gamma}{4\pi D (1-\exp\{-\gamma\,\tau( b^{i_1}+\cdots+b^{i_{\mitDelta s}})\})}}
	\nonumber
	\\
	&&\qquad\qquad
	\exp
	\left(-\frac{\gamma}{4D}
		\frac{z^2}{1-\exp\{-\gamma\,\tau( b^{i_1}+\cdots+b^{i_{\mitDelta s}})\}}\right),
	\label{P(z) Our Ehrenfest}
}%
and the variance as:
\eqn{
	& &
	{\rm Var}(Z_{\mitDelta t} (t))
	\nonumber
	\\
	&=&
	\frac{2D}{\gamma}
	\sum_{i_1,...,i_{\mitDelta s}} \frac{C^{\mitDelta s}}{Q^{i_1+\cdots+i_{\mitDelta s}}}
	(1-\exp\{-\gamma\,\tau( b^{i_1}+\cdots+b^{i_{\mitDelta s}})\}).
	\label{variance ELJ}
}%

When $ z $ is large, the density of ELJ in Eq. (\ref{P(z) Our Ehrenfest}) will be approximately given by:
\eq{
	P^{\rm ELJ}_{\tau\mitDelta s}(z)
	\simeq
	C' \sqrt{\frac{\gamma}{4\pi D}}\exp\left(-\frac{\gamma}{4D}z^2\right),
}%
with
\eq{
	C'\equiv
	\sum_{j_1,...,j_{\mitDelta s}} \frac{C^{\mitDelta s}}{Q^{j_1+\cdots+j_{\mitDelta s}}}
	\ (<1),
}%
where $ j_i $ of the summation is an integer satisfying the condition $ \gamma\,\tau( b^{j_1}+\cdots+b^{j_{\mitDelta s}}) \gg 1 $.
From the above equation, we find that the density of the proposed model decays in proportion to $ \exp (-z^2) $ toward the end of the tail as seen in Fig. \ref{density M Q2B3 with the lines estimated by log likelihood} (a), whereas in Fig. \ref{density M Q2B3 with the lines estimated by log likelihood} (b), the tail decays in proportion to $ z^{-\alpha-1} $ to mid-way when $ \mitDelta s $ is small.

The L\'evy stable symmetrical density function is given by:
\eqn{
	L(z,{\mitDelta} s \,|\,\alpha,\gamma) \equiv
	\frac{1}{\pi}\int_0^\infty \exp (-\gamma\,\mitDelta s\,q^\alpha)\cos(q\,z)\,dq,
	\label{Levy distribution}
}%
for the characteristic exponent $ \alpha $ ( $ 0<\alpha\le 2 $ ) and positive scale factor $ \gamma $.
This reduces to a Gaussian density function when $ \alpha $ = 2.
The L\'evy stable density is characterized by a fat tail and kurtosis, is consistent with empirical data, and satisfies a scaling law.
It is one of the most natural generalization of the Gaussian distribution which is stable with respect to convolution.

From Eq. (\ref{P(z) Our Ehrenfest}), we find the reason that the stable density function $ \alpha <2 $ is chosen for the best fitting density function of the proposed model for shorter $ \mitDelta s $ by the maximum likelihood method that we will explain later.
The large coefficient of $ z^2 $ in the exponential function gives a sharper peak at the origin, in contrast, the small coefficient of $ z^2 $ affords a flatter peak, broadening the density function.
Therefore, the density function given by the sum of these two types of terms with dissimilar coefficients has a fatter tail than the Gaussian density function.

When $ \mitDelta s $ is sufficiently large, the density function on proposed model becomes the sum of the Gaussian density functions of almost the same shape.
Therefore, the density function in the proposed model is given by the Gaussian density function for longer sampling intervals.

The proposed model well describes the changes in the stock indices such as the S\&P500, TOPIX and currency exchange rates.
Let each particle in the box $ + $ represent a buy stance dealer and each particle in the box $ - $ represent a sell stance dealer.
These dealers will get a lot of information from a newspaper, television, radio, business reports etc. on business conditions, weather reports, governmental decisions etc. and their view will be changed from buy to sell or from sell to buy on the basis of such information.
These changes will generate a change in the stock price or the exchange rate.
We assume that a change in the dealer's mind directly causes a change in the price so that $ M(s) $ is interpreted as the price.

As the information is spatially and temporally distributed, and any piece of information may have a different impact on the dealers, the price will sometimes change drastically within minutes, or sometimes only slightly over hours.
The difference between the progress of time in the ELJ and in the original Ehrenfest model describes this phenomenon.
The parameters $ b $ and $ Q $ might thus represent the dealers' (price change) sensitivity and frequency of information.

\section{Maximum likelihood method}

The maximum likelihood method \cite{Kitagawa} is employed to estimate parameters of the model.
It is assume that a joint density function $ f(x_1,...,x_n|\theta) $ of
the random variable $ (X_1,...,X_n) $ is given, where the parameter $ \theta $ specifies the density function.
If the random variables are mutually independent,
then the joint probability density function of $ (X_1,...,X_n) $ is given
as the product of the individual density functions of $ X_i,\ (i=1,2,...,n) $.
Thus we have
\eq{
	f(x_1,...,x_n|\theta)
	=
	f(x_1|\theta)f(x_2|\theta)\cdots f(x_n|\theta).
}%
Taking the logarithm on the right-hand side yields the log likelihood:
\eq{
	l(\theta)=\sum_{i=1}^n \log f(x_i|\theta).
}%
For the present case, the log likelihood is expressed as:
\eq{
	l(\alpha, \gamma) = \sum_s \log L(Z_{\mitDelta s}(s),\mitDelta s\,|\,\alpha,\gamma).
}%
The maximum likelihood estimate of the parameters of the model is obtained by maximizing this quantity.

\section{Numerical simulation}

The results of a numerical simulation of the trajectory of $ M(s) $ is shown in Fig.~\ref{M,Q2B3}, where $ N = 10000 $, $ b=3 $, and $ Q=2 $.
Big jumps that do not exist in the original Ehrenfest model are observed in the proposed model.
Because of these big jumps, the empirical density $ P_{\mitDelta s}(z) $ of the new model is a better fit to a stable distribution ($ \alpha < 2 $), whereas the empirical density $ P_{\mitDelta s}(z) $ is given by a Gaussian distribution in the original Ehrenfest model.

\begin{figure}
	\begin{center}
	\rotatebox{90}{\hspace{2.0cm}$ M(s)-N/2 $ }
	\includegraphics[height=6cm]{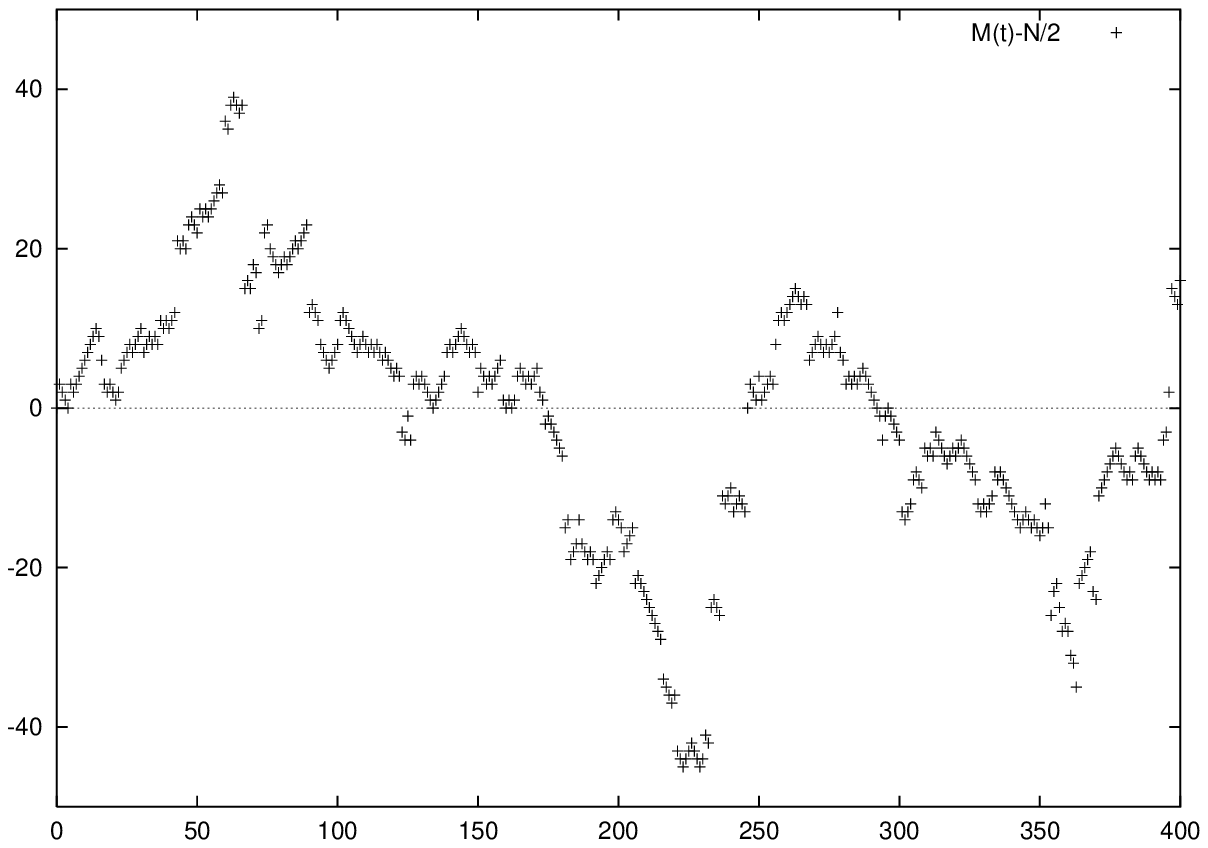}\\
	\qquad $ s $
	\caption[]
	{Trajectory of the number of particles in box $ + $ ($ M(s) $) for $ N = 10000 $, $ b=3 $, $ Q=2 $.}
	\label{M,Q2B3}
	\end{center}
\end{figure}

\begin{figure}
(a)
	\begin{center}
	\rotatebox{90}{\hspace{3cm}$ P_{\mitDelta s}(z) $ }
	\includegraphics[height=6cm]{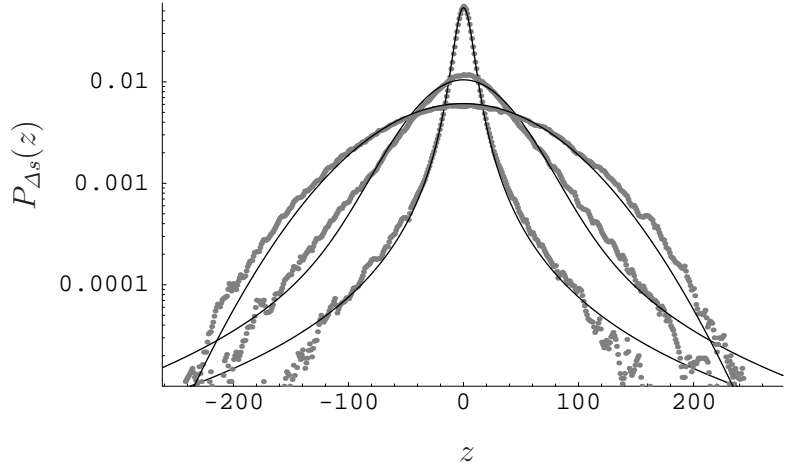}\\
	\vspace{-15pt}\qquad\qquad \, $ z $
	\end{center}
(b)
	\begin{center}
	\rotatebox{90}{\hspace{3cm}$ P_{\mitDelta s}(z) $ }
	\includegraphics[height=6cm]{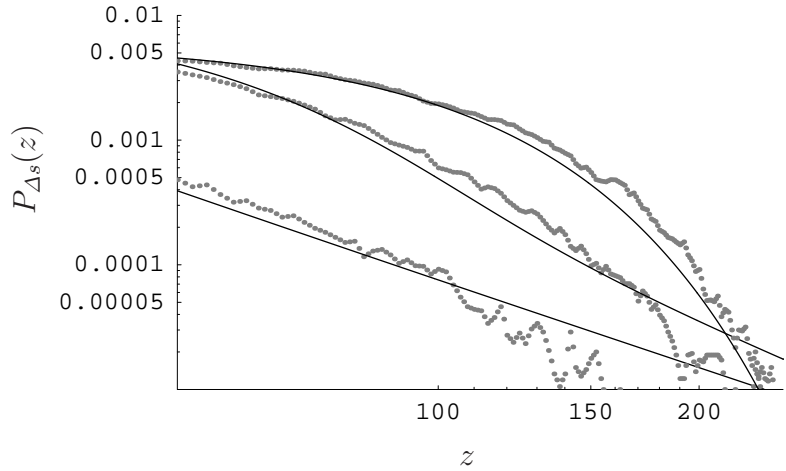}\\
	\vspace{-15pt}\qquad\qquad \, $ z $
	\end{center}
	\caption[]
	{Empirical densities $ P_{\mitDelta s}(z) $ for
	$ {\mitDelta} s = 8,\ 64,\ 512 $ (dotted line).
	$ N = 10000 $, and $ b=3,\ Q=2 $.
	The L\'vy stable density estimated using the maximum log likelihood method is also shown (solid line).
	$ \alpha = 1.310 $, 1.763, 2.0 and $ \gamma \mitDelta s = 9.223 $, 335.0, 2418.
	(a) Semi-log plot, (b) log-log plot.}
	\label{density M Q2B3 with the lines estimated by log likelihood}
\end{figure}

In Fig.~\ref{density M Q2B3 with the lines estimated by log likelihood}, the empirical density $ P_{\mitDelta s}(z) $ is shown for different time intervals $ {\mitDelta} s = 16,$ 64, 512.
These results were obtained for a sequence of about one million steps of the ELJ.
The density obtained by simulation using the proposed model is well described by a symmetrical L\'evy stable density function (\ref{Levy distribution}).

The parameters $ \alpha $ and $ \gamma \mitDelta s $ of Eq. (\ref{Levy distribution}) estimated using the maximum likelihood method for different time intervals are shown in Table \ref{table alpha, gamma}.
In the model, $ \alpha $ increase to $ 2.0 $ with increasing $ {\mitDelta}s $ for different values of $ N,\ Q $ and $ b $, as can be clearly understood from Table \ref{table alpha, gamma}.
Here we used the results of numerical simulation and considered the numerical values for $ \alpha $ and $ \gamma\mitDelta t $ for several values on $ N $, $ b $ or $ Q $ in Table \ref{table alpha, gamma}.
However, the probability density function of the proposed model can be approximately given by Eq. (\ref{P(z) Our Ehrenfest}).

\begin{table}
	\caption[]{$ \alpha $ and $ \gamma\mitDelta s $ estimated by the maximum likelihood method for $ a = 1 $.}
\begin{center}
\vspace{2pt}
	\begin{tabular}{l|cccccccc}
	\hline \hline
	& $ \mitDelta s $ & 16 & 32 & 64 & 128 & 256 & 512 & 1024 \\
	\hline
	$ N=10000 $ & $ \alpha $ & 1.386 & 1.529 & 1.763 & 1.958 & 2.0 & 2.0 & 2.0 \\
	$ b=3,\ Q=2 $ & $ \gamma \mitDelta s $ & 22.81 & 72.81 & 335.0 & 1285 & 2138 & 2418 & 2528 \\
	\hline
	$ N=1000 $ & $ \alpha $ & 1.758 & 1.956 & 2.0 & 2.0 & 2.0 & 2.0 & 2.0 \\
	$ b=3,\ Q=2 $ & $ \gamma \mitDelta s $ & 45.69 & 139.2 & 219.8 & 247.1 & 249.7 & 250.3 & 350.2 \\
	\hline
	$ N=10000 $ & $ \alpha $ & 1.719 & 1.756 & 1.804 & 1.859 & 1.932 & 1.988 & 2.0 \\
	$ b=2,\ Q=2 $ & $ \gamma \mitDelta s $ & 18.05 & 39.28 & 90.47 & 213.6 & 524.9 & 1155 & 1825 \\
	\hline
	$ N=10000 $ & $ \alpha $ & 1.713 & 1.748 & 1.796 & 1.871 & 1.949 & 2.0 & 2.0 \\
	$ b=3,\ Q=3 $ & $ \gamma \mitDelta s $ & 15.80 & 34.18 & 7813 & 196.6 & 493.7 & 1080 & 1685 \\
	\hline \hline
	\end{tabular}
\end{center}
	\label{table alpha, gamma}
\end{table}

Consider a density function given by a combination of an infinite number of Gaussian density functions:
\eqn{
	P^{\rm G}(x,\,\tau)
	=\frac{Q-1}{Q}
	\sum_j\frac{1}{Q^j}\sqrt{\frac{1}{2\,\pi\,\Delta^2\,b^j}}
	\exp\left(-\frac{x^2}{2\,\Delta^2\,b^j}\right).
	\label{eq. Brownian + Weierstrass}
}%
It is easy to confirm from the knowledge of domains of attraction \cite{Feller} that the density function of the above equation belongs to the domain of attraction of the symmetrical L\'evy stable density with characteristic exponent given by:
\eqn{
	\label{alpha = 2frac{log Q}{log b}}
	\alpha = 2\frac{\log Q}{\log b}.
}%
The density functions given by Eqs. (\ref{P(z) Our Ehrenfest}) and (\ref{eq. Brownian + Weierstrass}) are identical under a first order approximation in the limit $ \gamma \to 0 $ and $ \mitDelta s = 1 $.
Therefore, the symmetrical L\'evy stable density with characteristic exponent given by Eq.~(\ref{alpha = 2frac{log Q}{log b}}) fits the density function (\ref{P(z) Our Ehrenfest}) well as far as the central part of the density function of the ELJ is concerned.

\section{Exchange rate}

In ordered to discuss the empirical density function, we should redefine the empirical density function as a continuance version.
Price changes (returns) are defined by:
\eq{
	Z_{\mitDelta t}(t)\equiv S(t)-S(t-\mitDelta t),
}%
where $ S(t) $ is the price at time $ t $ and $ \mitDelta t $ is a time interval.
The empirical density function $ P_{\mitDelta t}(z) $ is then defined by:
\eq{
	\int_{z_1}^{z_2} P_{\mitDelta t}(z)\, dz =
	\int_{z_1}^{z_2} 
	\frac{\displaystyle\sum_t \delta (z-Z_{\mitDelta t}(t))}{n_{\rm last}}\,dz,
}%
where $ \delta (z) $ is the Dirac delta function and $ t $ takes $ n \mitDelta t' $, $ n $ = 1, 2,...,$ n_{\rm last} $.

\begin{figure}
	\begin{center}
	\rotatebox{90}{\hspace{3cm}$ P_{\mitDelta t}(z) $ }
	\includegraphics[height=6cm]{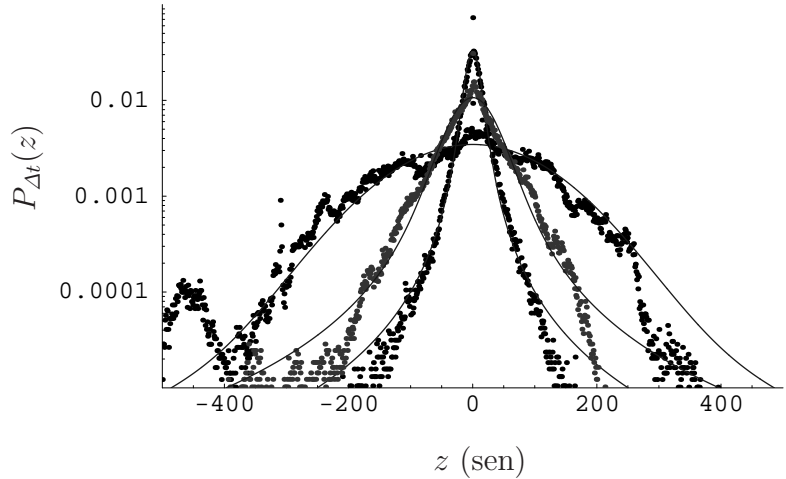}\\
	\vspace{-12pt}\qquad~~~~~~~~~ $ z $ (sen)
	\end{center}
	\caption[]
	{Empirical density functions $ P_{\mitDelta t}(z) $ for sampling intervals of
	$ {\mitDelta} t = 2^{15}$, $ 2^{18}$, $ 2^{21}$ seconds (dots) and the corresponding stable density functions (lines,
	$ \alpha = $ 1.451, 1.588, 1.900 and
	$ \gamma \mitDelta t $ = 23.86, 184.5, 4276)
	estimated by the maximum log likelihood.
	}
	\label{kawa94uy}

\end{figure}

The empirical density of the exchange rate between the US dollar and the Japanese yen over one year (Jan. '94-Dec. '94) \cite{Olsen and associate} is shown in Fig. \ref{kawa94uy}, along with the estimated stable density functions.
This figure shows that the empirical density is close to the stable distribution with a characteristic exponent $ \alpha < 2 $ for short sampling intervals and a Gaussian distribution for long sampling intervals.
Table \ref{table kawase94} shows the maximum log likelihood estimate of $ \alpha $ and $ \gamma \mitDelta t $.
When the sampling interval is small, the return is distributed according to a stable law with characteristic exponent $ \alpha < 2 $.
In addition, $ \alpha $ becomes larger with expansion of the sampling interval and approaches 2.
Similar results were obtained for German stocks \cite{Eberlein 1995}.

\begin{table}
	\caption[]{ $ \alpha $ and $ \gamma\mitDelta t $ estimated by the maximum log likelihood for the empirical density of the exchange rate \cite{Olsen and associate}.}
\vspace{2pt}
\begin{center}
	\begin{tabular}{ccccccccc}
	\hline \hline
	$ \mitDelta t $ & $2^{14}$ & $2^{15}$ & $2^{16}$ & $2^{17}$ & $2^{18}$ & $2^{19}$ & $2^{20}$ & $2^{21}$ \\
	\hline
	$ \alpha $ & 1.436 & 1.451 & 1.441 & 1.483 & 1.588 & 1.666 & 1.840 & 1.900 \\
	\hline
	$ \gamma \mitDelta t $ & 13.84 & 24.86 & 38.95 & 75.46 & 184.5 & 431.75 & 1743 & 4276 \\
	\hline \hline
	\end{tabular}
\end{center}
	\label{table kawase94}
\end{table}

\section{Concluding remarks}

This paper introduced a simple stochastic model (ELJ) to understand that the observed density function of the price change is approximately given by a truncated symmetrical L\'evy stable distribution for short sampling intervals and a Gaussian distribution for long sampling intervals.

The elastic potential of the Ehrenfest model makes the tail of the density function decay rapider than that of the L\'evy stable density.
The empirical density function of the price change should not have an infinitely long heavy tail.
Thus, the empirical density function given by Eq. (\ref{P(z) Our Ehrenfest}) seems to be reasonable.

\vspace{20pt}
\noindent
{\large\bf Acknowledgement}
\vspace{10pt}

The author would like to thank Professors Y. Itoh, G. Kitagwa, T. Ozaki, M. Taiji and Y. Tamura for helpful suggestions and comments.

%

\end{document}